# Ultrafast Parallel LiDAR with Time-encoding and Spectral Scanning: Breaking the Time-of-flight Limit

Zihan Zang[1,†], Zhi Li[1,†], Yi Luo[2,3,4,*], Yanjun Han[2,3,4,*], Xuanyi Liu[1], & H.Y. Fu[1,*]

[1]Shenzhen International Graduate School (SIGS), Tsinghua University, Shenzhen 518055, China.
[2]Beijing National Research Center for Information Science and Technology (BNRist), Department of Electronic Engineering, Tsinghua University, Beijing, 100084, China.
[3]Flexible Intelligent Optoelectronic Device and Technology Center, Institute of Flexible Electronics Technology of THU, Zhejiang, Jiaxing, 314006, China.
[4]Center for Flexible Electronics Technology, Tsinghua University, Beijing, 100084, China.
[†]Equal contribution authors
[*]E-mail: luoy@tsinghua.edu.cn, hyfu@sz.tsinghua.edu.cn

(Date 3/9/2021)

## Abstract

Light detection and ranging (LiDAR) has been widely used in autonomous driving and large-scale manufacturing. Although state-of-the-art scanning LiDAR can perform long-range three-dimensional imaging, the frame rate is limited by both round-trip delay and the beam steering speed, hindering the development of high-speed autonomous vehicles. For hundred-meter level ranging applications, a several-time speedup is highly desirable. Here, we uniquely combine fiber-based encoders with wavelength-division multiplexing devices to implement all-optical time-encoding on the illumination light. Using this method, parallel detection and fast inertia-free spectral scanning can be achieved simultaneously with single-pixel detection. As a result, the frame rate of a scanning LiDAR can be multiplied with scalability. We demonstrate a 4.4-fold speedup for a maximum 75-m detection range, compared with a time-of-flight-limited laser ranging system. This approach has the potential to improve the velocity of LiDAR-based autonomous vehicles to the regime of hundred kilometers per hour and open up a new paradigm for ultrafast-frame-rate LiDAR imaging.

# Introduction

To obtain the three-dimensional depth information of a long-range object, time-of-flight (TOF) imaging is one of the most powerful methods with high accuracy and high transverse spatial resolution. TOF imaging-based light detection and ranging (LiDAR) has been wieldy used in industrial automation[1], logistics[2], autonomous driving[3], robotics[4] and airborne remote sensing[5], etc. For these applications, the acquisition time is crucial, because fast-changing scenes and fast-moving objects require a high image refresh rate to perform data post-processing such as object identification, tracking and measurement.

The acquisition time of a scanning LiDAR is determined by two factors: the speed of beam scanning and the maximum time of flight. For most commercial LiDARs, the speed of beam scanning is one of the main bottlenecks. Mechanically scanning LiDARs suffer from slow scanning rate of several kHz. State-of-the-art optical phased array (OPA) technology has the potential of MHz-to-GHz scanning rate, but the large-scale integration of low-loss active optical antennas is difficult to achieve, and the high peak optical power requirement of TOF imaging is hard to fulfill with current photonic integration platforms. Spectral scanning using a spatially dispersive element and a broadband source is another promising method to achieve uniquely high speed, which is inherently inertia-free and compact. Recently, spectral beam scanning with near-THz scanning rate has been reported[6]. Microscopy[7,8], endoscopy[9], photography[10] and LiDAR[11,12] based on spectral scanning have been demonstrated with record frame rates.

As the pursuit of rapid beam-steering method pushes up the speed of current LiDAR, the time consumption of round-trip delay of light pulses becomes increasingly remarkable. Next-generation high-speed autonomous vehicles such as unmanned aerial vehicle (UAV) push or exceed the so-called time-of-flight limit. For example, to resolve a movement of 0.1 m, which is the baseline depth resolution of most TOF imaging systems, an autonomous vehicle with a speed of 100 km/h and at a distance of 100 m requires an acquisition rate higher than 5.6 MHz to generate a normal 3D image with 100 lines and 200 pixels per line. However, a TOF limited ranging system can only offer a detection rate up to 1.5 MHz for a detection range of 100 m. Thus, a few times of speed improvement beyond TOF limit is critical. TOF limit is mainly due to the serial detection mode of a scanning LiDAR and can be alleviated by introducing parallelism. The most direct approach to implement parallelism is the use of pixelated arrays, which can eliminate the need of beam scanning. However, high-resolution, long-range LiDAR based on TOF imaging preferred by industrial automation and autonomous driving requires high-speed and highly sensitive photodetectors with bandwidth of tens of GHz and corresponding high-speed sampling circuits. However, it is difficult to realize large-scale integration for such detector arrays. Recently, parallel coherent LiDAR was demonstrated using a soliton microcomb[12]. With heterodyne detection, detectors with relatively low bandwidth can be applied. Nevertheless, to obtain high spatial resolution, large-scale integration of many high-performance detectors is still needed. Therefore, it is highly desirable to find a method to implement parallelism for LiDARs using a single-pixel detector. Compressive sensing-based LiDAR with a pixelated spatial modulator can achieve parallel detection with a single detector[13,14,15]. However, the need of multiple measurements, high computational complexity and the speed of spatial modulators slow down the actual speed. Temporal encoding is another method to implement parallelism with a single receiver. Temporal encoded imaging systems based on electro-optic modulation[17-19] or optical feedback[20] enables parallelism and thus reduce the acquisition time, but the speed limitations caused by mechanical beam scanning devices and modulators make it impractical to surpass the time-of-flight-limited speed.

Here, we demonstrate a parallelism technique to enable scalable speed improvement beyond time-of-flight limit for a single-pixel scanning LiDAR, and fast spectral scanning is incorporated to offer an unlimited inertia-free beam scanning. As shown in Fig. 1a, by employing a fiber-based all-optical temporal encoder in each wavelength multiplexed channel, optical code division multiplexing (OCDM) is implemented to achieve parallelism. Key to our approach is the generation and leverage of a correlated spectro-temporal modulation with high degree of freedom (DOF). In comparison, previous spectro-temporal encoded imaging systems[10,16] use a serial time-encoding strategy (Figs. 1b and 1f). As the dispersive elements perform a one-to-one spectrum-to-space mapping, only serial detection can be achieved, and the imaging speed is TOF limited. Using the proposed approach, we exploit correlations between the spectral and temporal DOFs to enable parallel TOF ranging for the first time, leading to multiplication of detection speed. It is also worth noting that high-DOF spectro-temporal modulation has recently been employed in parallel optical computing[21,22]. The speed multiplier effect of our proposed system relies on the temporal code size. A set of temporal code with $N$ types of code words can generate $M$ parallel channels and thus reduces the acquisition time of a frame (Figs. 1b-1e). Note that the final multiplier is slightly less than *M* because of the additional cost of encoding time. Moreover, benefiting from programmable all-optical fiber-based encoders[23], the scale of the speed multiplication is reconfigurable. A partially correlated spectro-temporal encoder (Fig. 1g) with a simpler structure offers a smaller multiplier (Figs. 1c and 1d) while a fully correlated spectro-temporal encoder (Fig. 1h) enables single-shot detection (Fig. 1e) at the cost of complexity. By implementing partially correlated spectro-temporal encoders on 45 distinct wavelength channels, we generate 5 parallel OCDM channels with optical orthogonal code (OOC)[24-26], thus reducing the number of measurements from 45 to 8. For ranging distance of 75 and 25 meters, we achieve 4.4 and 3.6 times the time-of-flight-limited speed, respectively. The proposed parallelism approach can enable ultrafast long-range scanning LiDAR much faster than the time-of-flight-limited speed, which is highly desirable for the aforementioned applications.

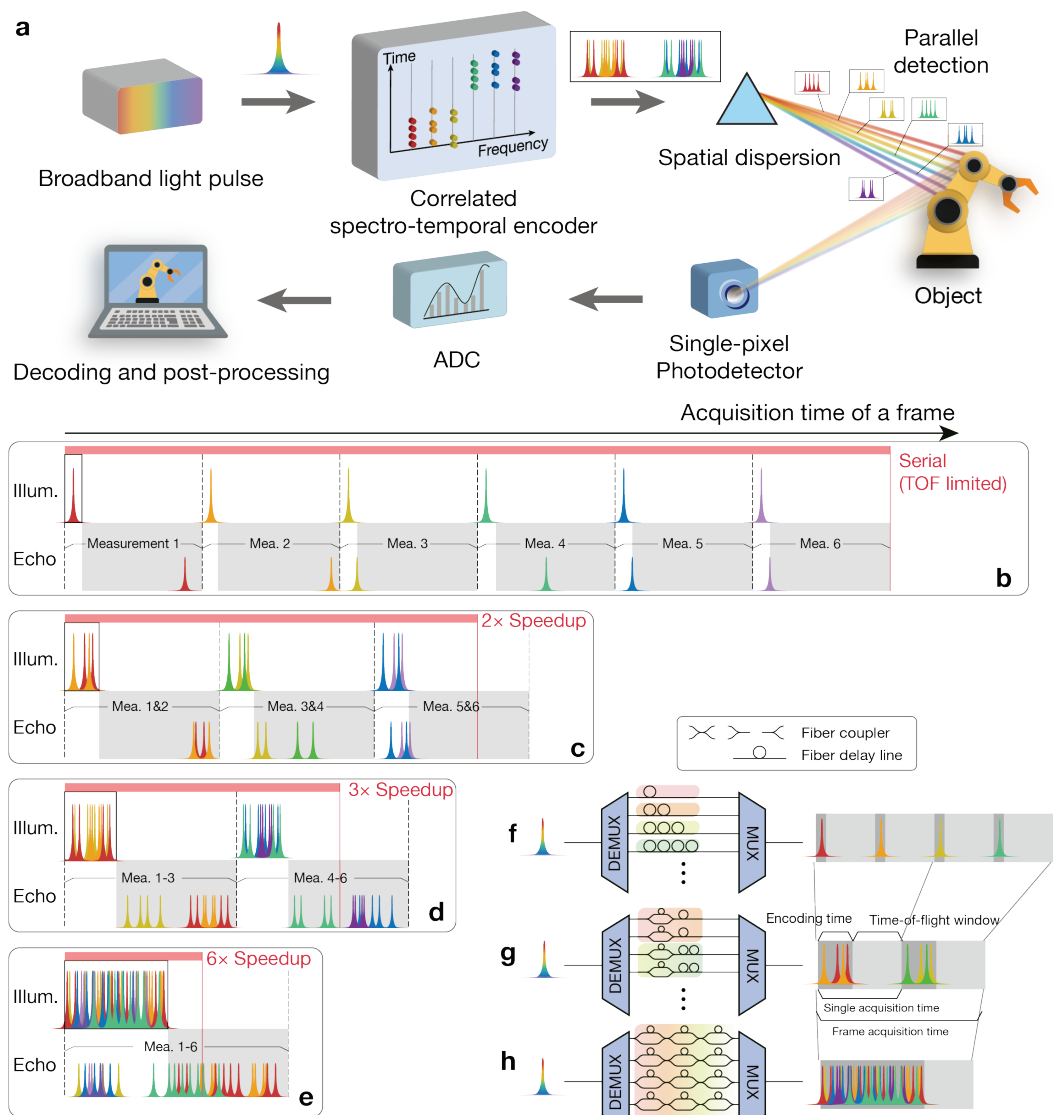

**Figure 1 | All-optical correlated spectro-temporal encoding enables scalable speedup of LiDAR beyond time-of flight limit. a**, a correlated spectro-temporal encoder modulates the input broadband light pulse. A dispersive element performs the spectrum-to-space mapping and thus achieve spectral beam scanning. Meanwhile, parallel ranging by a single-pixel detector is achieved by temporal encoding and decoding. **b**, illumination and echo signal of a serial spectro-temporal encoded LiDAR, which is time-of-flight limited. Because of the lack of parallelism, 6 scan periods are required for a frame with 6 ranging points. **c-e**, illumination and echo signals of correlated spectro-temporal encoded LiDAR proposed here. Spectro-temporal encoding with different degree of correlation enables parallelism and speeds up the LiDAR beyond time-of-flight limit. The final speedup multiplier is slightly less than parallelism multiplier because of the additional cost of encoding time. **f**, a serial spectro-temporal encoder used by **b**, where different delays are introduced for different spectral channels. **g**, a partially correlated spectro-temporal encoder used by **c** and **d**, where spectral channels are divided into several groups, and different groups share the same set of optical orthogonal code but different global delay. The channels within a group will operate parallelly. **h**, a fully correlated spectro-temporal encoder used by **e**, where a set of optical orthogonal code is applied to all spectral channels, and all channels will operate in parallel.

# Results

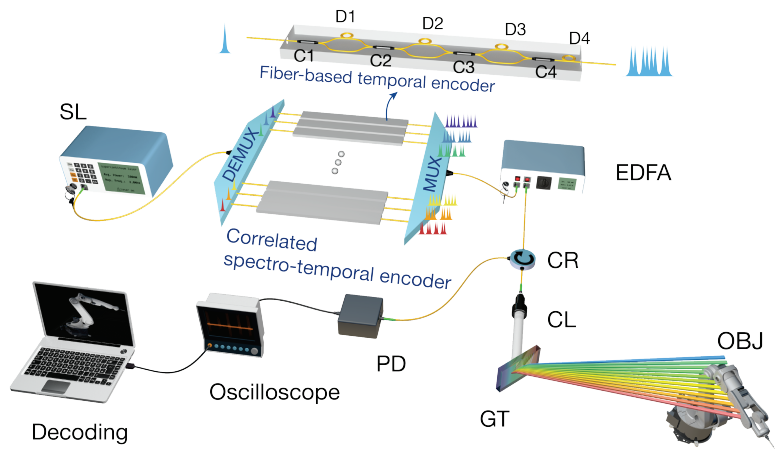

**Figure 2 | Experimental setup.** A wavelength division demultiplexer (DEMUX) splits the input broadband light pulse generated by a supercontinuum laser (SL) into multiple spectral channels. A fiber-based encoder array performs temporal encoding for each channel, and all the channels are combined by a wavelength division multiplexer (MUX). A correlated spectro-temporal encoded light pulse train is thus generated and transmitted

for spectral scanning and time-of-flight ranging. A diffraction grating (GT) performs horizontal spectral scanning; vertical scanning is implemented mechanically using a rotatable mirror (not shown in figure). The echo light from the object is received by a photodetector and the temporal encoding is sampled and decoded by post-processing. C1-C4, fiber couplers; D1-D4, fiber delay lines; EDFA, Erbium-Doped Fiber Amplifier; CR, optical circulator; CL, collimator; PD, photodetector; ADC, analog-to-digital convertor.

**Experimental setup:** As shown in Fig. 2, a gain-switched supercontinuum laser with an optical filter covering 1540-1560 nm provides broadband light pulses with a width of 100 ps and variable repetition frequency. Each light pulse is spectro-temporally modulated with high DOF by a fiber-optic encoder, which will be described below. The modulated pulse is then amplified by an erbium-doped fiber amplifier (EDFA) and transmitted through a collimator. Spectral scanning is achieved by a diffraction grating. The echo light is collected by the same optics, redirected by an optical circulator and received by a high-speed APD (1.5-GHz bandwidth, linear mode). The echo signal from the detector is sampled and captured by a high-speed oscilloscope (Tektronix MSO72304DX, 33GHz) and decoded via post-processing hardware.

**Correlated spectro-temporal encoding:** The correlated spectro-temporal encoder (see Fig. 2) is comprised of two arrayed waveguide gratings (AWGs), and a series of homemade fiber-based temporal encoders. AWGs are used as wavelength multiplexer and demultiplexer to generate 45 distinct spectral channels. Considering the complexity of the fiber-based temporal encoder, a partially correlated spectro-temporal encoder similar to Fig. 1g is chosen, where 45 spectral channels are divided into 8 groups. Each group contains 5 parallel OCDM channels, each of which contains an encoder with 4 fiber couplers (C1-C4 in Fig. 2) and 3 fiber delay lines (D1-D3 in Fig. 2). The encoder is able to generate 5 kinds of OOCs by adjusting the delays. All generated OOCs has a whole code length (length of the binary sequence) of 56, which contains 8 non-zero bits. The time slot (the minimum encoding temporal interval) is chosen as 1.2 ns, according to the available bandwidth of the photodetector. Therefore, the total encoding time $t_{code}$ of the OOC is 67.2 ns. The encoding result of one of the parallel groups is shown in Fig. 3a. Other details of the construction of OOCs can be found in the Methods and Supplementary Information. As each group shares the same set of OOCs, incremental global time delays (D4 in Fig. 2) are introduced for different groups to separate them in the time domain, which is similar to the time-stretch technique[11]. The increment of the global delay $T_d$ corresponds to a single acquisition cycle, which equals to the sum of the encoding time $t_c$ and the time-of-flight window $\tau$ for the maximum range without ambiguity (see Figs. 1f-h). $T_d$ is set to be 167 ns for 25-m maximum detection range and 500 ns for 75-m case, respectively. Thus, for a frame with $N$ effective ranging pixels, $M$ parallel channels for each group, and thus $N/M$ groups, the total acquisition time $T$ for a frame is given by $N(t_c+\tau)/M$, which is also the repetition period that the laser source should be configured. Compared with a serially scanning LiDAR with a total acquisition time $N\tau$, the multiplier of speed improvement is $M/(1+t_c/\tau)$, which increasingly approaches $M$ as $t_c/\tau$ decreases, i.e., the encoding time becomes unremarkable compared with the time-of-flight window.

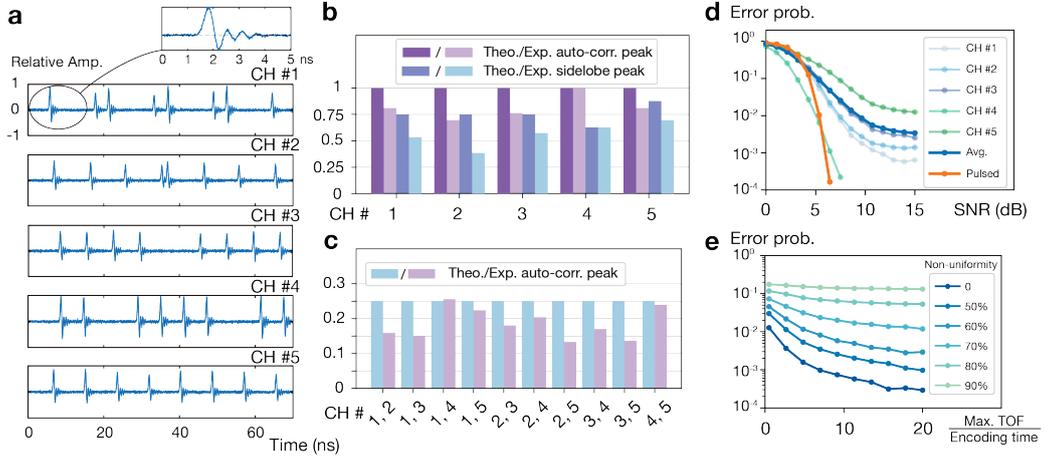

**Figure 3 | Encoding and decoding performance. a**, encoding waveform of 5 OCDM channels belonging to one of the groups, captured by a photodetector with 1.5-GHz bandwidth; a single pulse is zoomed in. **b**, theoretical and experimental results of normalized auto-correlation main peak value $\lambda_p$ and maximum auto-correlation sidelobe peak $\lambda_a$ of each OCDM channel. **c**, theoretical and experimental results of normalized cross-correlation peak value $\lambda_c$ between each two different OCDM channels. **d**, error probability under different signal-to-noise ratio, estimated by Monte Carlo analysis with $5\times10^4$ simulations for each point. The results of 5 OCDM channels and the averaged result are shown. For comparison, error probability of a pulsed LiDAR is also provided. **e**, error probability under different detection time-of-flight window (i.e., maximum time of flight, normalized by the encoding time) and intensity non-uniformity among different channels (without additional noise, and thus showing the impact of MAI), estimated by Monte Carlo analysis with $5\times10^4$ simulations for each point.

**Temporal decoding results:** It is possible that the echo light belonging to the same parallel group arrive at the same time or overlap with each other, and the arriving order is also uncertain. To identify the exact time of flight of a specific channel, temporal decoding is needed. Temporal decoding is achieved by correlating the sampled echo signal $e(t)$ with each OOC codeword $c_i(t)$, i.e. $\langle e, c_i \rangle = \int e(t' - t) c_i(t') \mathrm{d}t'$, which can be calculated efficiently and parallelly with current hardware[27,28]. Due to the orthogonality of OOCs, the auto-correlation signal exhibits a high peak value, whose position corresponding to the arriving time of the codeword, while the cross-correlation signals corresponding to other OCDM channels show low peak values. Cross-correlation signals and the sidelobe of auto-correlation signal would lead to decoding noise, which is denoted as multiple access interference (MAI). An ideal OOC set $\{c_i\}$ with low-level MAI consists of white-noise-like codewords with good orthogonality and randomness. Orthogonality can be evaluated by the maximum value $\lambda_c$ of the cross-correlation between different codewords $\langle c_i, c_j \rangle$ $(i \neq j)$, which approaches zero in the ideal case. Randomness can be evaluated by the maximum value $\lambda_a$ of auto-correlation sidelobes of each codeword $\langle c_i, c_i \rangle$, which also approaches zero in the ideal case. Normalized $\lambda_c$ and $\lambda_a$ of one of the OCDM group calculated both from codebook and experimentally captured waveform are shown in Figs. 3b and 3c. The height of the normalized auto-correlation main peak $\lambda_p$ is also depicted in Fig. 3b, which reflects the relative intensity differences among OCDM channels. The difference between theoretical and experimental results is mainly due to the fiber length error for fixed fiber delay lines and slight differences in the loss and coupling of the fiber-optic devices.

To assess the impact of MAI, error probability of this OCDM LiDAR is estimated by Monte Carlo simulation, using the experimentally captured waveforms. Figure 3d shows the error probability under different noise level. Compared to conventional pulsed LiDARs, the total noise of this LiDAR includes not only the detection noise but also the MAI. As a result, the signal-to-noise ratio (SNR) requirement for high-fidelity detection is higher for an OCDM LiDAR. Figure 3e shows the error probability induced by MAI (without additional noise). As MAI only exists when two or more codewords overlaps, which is called a collision, a longer detection temporal window (compared with the encoding time of a codeword) can alleviate the impact of MAI. On the other hand, intensity non-uniformity among different OCDM channels will exacerbate the MAI, because cross-correlation peaks formed by a strong channel may exceed the auto-correlation peak from a weak channel. In addition, the OOC coding selection also determines the magnitude of the MAI. In other words, the cost of parallelism here achieved by OCDM is a slight reduction in the signal-to-noise ratio caused by MAI.

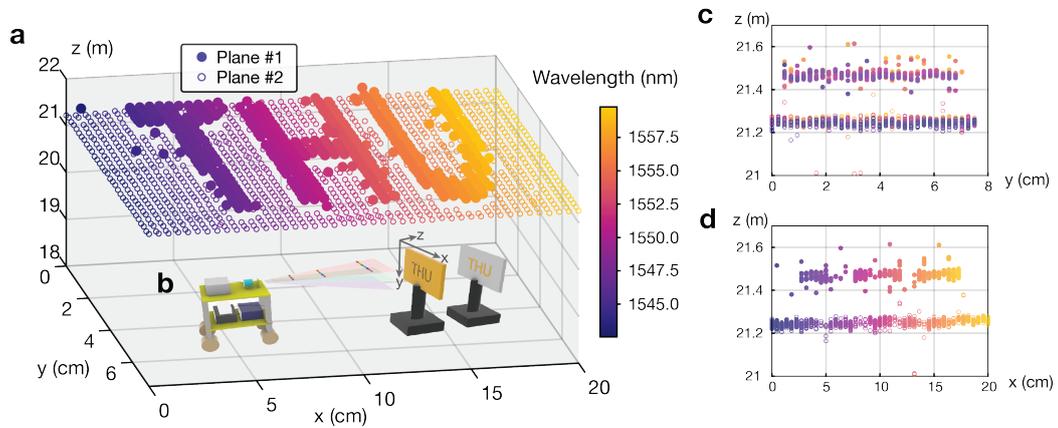

**Figure 4 | 3D Imaging result. a**, 3D image of the two targets with Tsinghua University logo (THU). The two targets are positioned 6.5 m away. The result contains a 14.8-m effective length of fiber link. Spectral scanning generates 45 channels along horizontal axis (y). Vertical scan is performed by a rotatable mirror. Large circles indicate the front target (plane #1) and small circles indicate the back target (plane #2). **b**, experimental set-up. **c**, projection of the 3D image along the *x* axis. **d**, projection of the 3D image along the *y* axis.

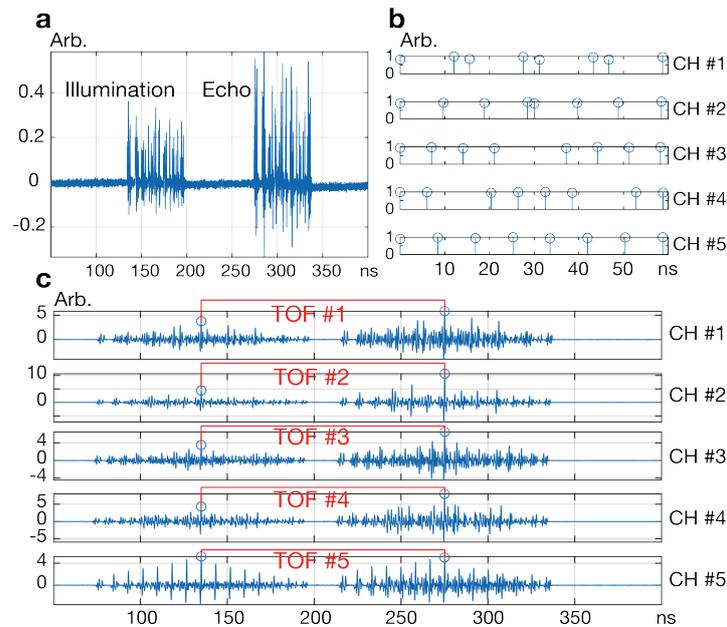

**Figure 5 | Decoding result. a**, sampled echo signal of a parallel group including 5 OCDM channels. **b**, temporal waveform of the 5 OOC codewords. Intensity fluctuations due to fiber optic encoders are incorporated. **c**, after correlating the echo signal with the 5 OOCs, 5 decoded signals are obtained. Two marked auto-correlation peaks in each decoded channel indicate the transmission time of detection light and the arriving time of echo, respectively.

**Imaging performance:** To demonstrate 3D imaging performance of the proposed LiDAR, targets are set as two pieces of cardboards spaced by 0.5 m, one of which is hollowed out in the shape of Tsinghua University logo (see Fig. 4b). 45 spectral channels spanning from 1540 nm to 1560 nm are dispersed by a diffraction grating with a line density of 800 line/mm, realizing horizontal line scan with a field of view (FOV) of 2° and covering the logo with a width of 20 cm. The vertical scan is performed mechanically using a rotatable mirror. The maximum detection range is chosen as 25 m, corresponding to a maximum detection rate of 6 MHz for a time-of-flight limited serially scanning LiDAR. By using the 5-channel OCDM, a maximum detection rate of 21.8 MHz is achieved, leading to 3.6-fold speedup beyond time-of-flight limit. The sampled echo signal of a parallel 5-channel group is shown in Fig. 6a. After correlated with 5 OOCs shown in Fig. 6b, the decoded signals of 5 channels are obtained, as illustrated in Fig. 6c. Despite the appearance of MAI sidelobes, two main auto-correlation peak stands out prominently in each decoded signal, which indicates the arriving time of illumination and echo signal belonging to each channel, respectively. As a result, the time-of-flight results of the 5 parallel channels can be obtained during a single detection cycle. The imaging result is depicted in Figs. 4a, 4c and 4d, in which both target planes are clearly captured.

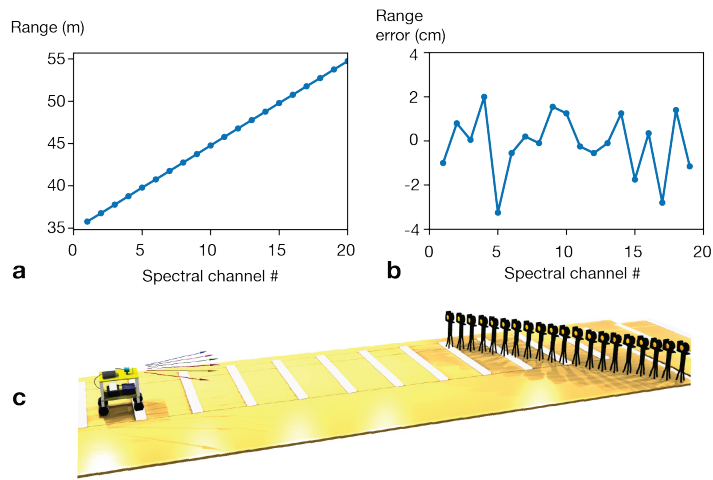

**Figure 6 | Long-range detection result. a**, ranging result of 20 spectral channels. The targets are 20 poles with reflective tags and placed at a range from 21 m to 40 m. The result contains a 14.8-m effective length of fiber link. They are equidistant from each other at a distance of 1 m. **b**, ranging error of the distance between each adjacent pole, which exhibits a depth resolution of several centimeters. **c**, experimental setup.

**Long-range performance:** One advantage of the parallel LiDAR is the ability to perform high-speed long-range imaging. We constructed a long-range 3D scene containing 20 poles with reflective tags (see Fig. 6c). Only horizontal spectral scanning is performed, where 20 out of 45 spectral channels are selected and divided into 4 groups, each containing 5 OCDM parallel channels. Another diffraction grating with a line density of 1200 line/mm is used to achieve a FOV of 3.8°. The maximum detection range is set as 75 m, corresponding to a maximum detection rate of 2 MHz for a time-of-flight limited serially scanning LiDAR. With parallelism, detection rate is increased to 8.8 MHz, indicating 4.4 times the time-of-flight limited speed. The accuracy performance is depicted in Figs. 6b and 6c, confirming accurate range mapping even for relatively long-range situation.

## Discussion

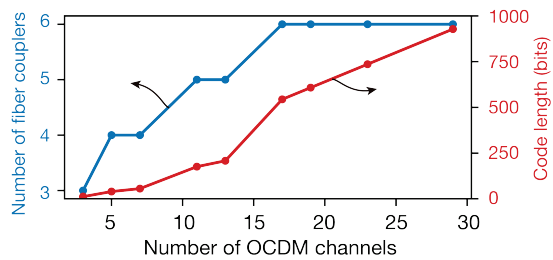

**Figure 7 | Scalability of OCDM channels using quasi-prime code.** As the number of OCDM channels increases, the scale of parallelism increases but code length and the required number of optical couplers per channel increases as well.

The speedup scalability of the proposed LiDAR mainly depends on the number of channels $M$ within an OCDM group. However, large parallel channels $M$ has two main drawbacks, i.e., the increase of code length and the number of required optical couplers in fiber-optic encoder, which is depicted in Fig. 7 for the case of quasi-prime code. The increase of code length or encoding time will reduce actual multiplier of speed improvement because of the time consumption of code transmission, and the large number of fiber couplers will increase system complexity. To address the limitation of scalability, it is possible to reduce the encoding time by implementing a smaller time slot. In our experiment, the minimum time slot is limited by the 1.5-GHz detector bandwidth. If a detector with higher speed is used, such as a 40-GHz detector, the encoding time can be largely reduced and thus leads to better speedup effectiveness. Benefiting from the single-pixel architecture in this system, such high-speed detectors are practical to employ. Furthermore, if high-speed electro-optical modulators are used in place of fiber-optic encoders for OCDM code generation, OOCs other than quasi-prime code and with shorter code length can be adopted, which not only extends the scalability of the system, but also helps improve the system performance due to the reduced MAI.

Another potential issue of the proposed system is the reduced peak illumination power caused by temporal encoding. This effect becomes more notable as the number of non-zero bits in codeword increases, which also limits the scalability of the system. When code weight (the number of non-zero bits) equals to $Q$, the peak power becomes $1/Q$ of the unmodulated case. For the case of quasi-prime code, $Q$ equals to $2^L$ where $L$ is the number of fiber couplers, and thus the impact on scalability can also be seen in Fig. 7. During the correlation process, signal peaks corresponding to non-zero bits are constructively added back to the unmodulated case, whereas thermal noise (assumed to be Gaussian) is incoherently added and become $\sqrt{Q}$ times of the unmodulated case. As a result, effective illumination power reduction of the system is $1/\sqrt{Q}$ when compared to a single-pulsed LiDAR. Nevertheless, this reduction effect does not take into account the presence of subsequent EDFAs. The insertion of a simple pre-amplifier will effectively compensate the loss of illumination power[29]. Besides, as the fiber-based temporal encoder only replicates input pulses according to quasi-prime code (see Methods), the generated pulses with nearly identical pulse width with input pulses are relatively evenly distributed, which is beneficial to pulse amplification.

In conclusion, we propose and demonstrate a method to implement parallelism for single-pixel scanning LiDARs. Scalable parallelism and rapid spectral scanning are achieved simultaneously by correlated spectro-temporal modulation using all-optical fiber-optic encoders and a dispersive element. A line scan rate of several hundreds of kilohertz can be achieved even for tens of meters of detection range, breaking the time-of-flight limit of serially scanning LiDARs. Our approach can enable next-generation high-speed autonomous vehicles and manufacturing with moving speed beyond hundred kilometers per hour. Also, this method can be further extended to other active imaging systems which uses serially scanning illumination, such as multiphoton microscopy[8] and ultrafast imaging[10,16].

## Methods

**OCDM encoding:** Quasi-prime code (QPC)[25] is employed to implement all-optical OCDM. The generation of QPC codeword set is as follows.

Firstly, choose a prime number $P$, and construct $P$ sequences $\{S_i^P, i = 0 \text{ to } P − 1\}$, each with $P$ elements. The $j$-th element of the $i$-th sequence is then given by

$$S_i^P(j) = (i \cdot j) \bmod P, \text{ for } i = 0 \text{ to } P − 1; j = 0 \text{ to } P − 1.$$

Secondly, convert $\{S_i^P\}$ into $P$ binary sequences $\{C_i^P, i = 0 \text{ to } P − 1\}$, each with $P$ $P$-element frames, i.e. $P^2$-element sequence. Each frame contains only one non-zero bit. The binary sequences are obtained using the following rules:

$$C_i^P(n) = \begin{cases} 1, & \text{for } n = jP + S_i^P(j); j = 0 \text{ to } P − 1 \\ 0, & \text{otherwise.} \end{cases}$$

Thirdly, choose $Q$ as the smallest power of two that is larger than $P$, and construct a new set of $P^2$ sequences $\{C_{ik}^{QP}, i = 0 \text{ to } P − 1; k = 0 \text{ to } P − 1\}$, each with $Q$ $P$-element frames, i.e. $QP$-element sequence. The n-th element of a sequence is given by

$$C_{ik}^{QP}(n) = C_i^P\big((n + kP) \bmod P^2\big), \text{ for } n = 0 \text{ to } QP − 1.$$

Finally, the QPC set is generated by discarding nonsymmetric code from the set $\{C_{ik}^{QP}\}$. Each final QPC codeword is symmetric, with a code length of $QP$ as well as code weight of $Q$.

**All-optical fiber-optic encoder:** An all-optical QPC encoder requires $\log_2 Q$ interferometers, which contains $\log_2 Q + 1$ fiber couplers and $\log_2 Q$ fixed fiber delay lines to achieve a QPC codeword with code weight of $Q$. The suitable delays can be easily found by checking non-zero bits in the codeword. The details of the delay lines can be found in the Supplementary Information.

# Acknowledgements

The authors would like to express sincere thanks to Tsinghua-Berkeley Shenzhen Institute (TBSI) Faculty Start-up Fund.

# Author contributions

Z. Z. and Z. L. proposed the idea, performed the experiments and designed the Monte Carlo simulations. Z. Z., Z. L. and X.L. performed system analysis and prepared the laser source. H.F., Y.H., and Y.L. contributed to the data analysis. Y.L. and H.F. supervised the research. All authors discussed the manuscript together.

# Competing interests

The authors declare no conflicts of interest.

# Data availability

The data that support the findings of this study are available from the corresponding author upon reasonable request.